\documentclass[11pt,a4paper]{article}

\usepackage{amsmath}
\usepackage{amsfonts}
\usepackage{amssymb}
\usepackage{graphicx}
\usepackage{hyperref}
\usepackage{float}
\usepackage{cite}
\usepackage{version}
\numberwithin{equation}{section} 
\usepackage{geometry}
\usepackage{color}
\author{\textsc{Solard} Gautier}
\newcommand{\be}{\begin{equation}}
\newcommand{\ee}{\end{equation}}
\newcommand{\beq}{\begin{equation}}
\newcommand{\eeq}{\end{equation}}
\newcommand{\ba}{\begin{eqnarray}}
\newcommand{\ea}{\end{eqnarray}}
\newcommand{\bea}{\begin{eqnarray}}
\newcommand{\eea}{\end{eqnarray}}

\renewcommand{\Re}{\operatorname{Re}}
\renewcommand{\Im}{\operatorname{Im}}
\begin{document}
\pagestyle{empty}
\begin{center}

\begin{Large}
\textbf{A note on the Ricci scalar of six dimensional manifold with SU(2) structure}
\end{Large}

\vspace{0.1cm}

Gautier Solard

\emph{Physics Department, Universit\`a  di Milano-Bicocca, Piazza della Scienza 3,20100 Milano, ITALY}

\emph{INFN, sezione di Milano-Bicocca, Milano, ITALY}

gautier.solard@mib.infn.it
\end{center}

\begin{abstract}
Taking \cite{2007JGP} as an inspiration, we study the intrinsic torsion of a SU(2) structure manifold in six dimensions to give a formula for the Ricci scalar in terms of torsion classes. The derivation is founded on the SU(3) result coming from the aforementioned paper.
\end{abstract}

\section*{Introduction}

Generalized geometry \cite{Hitchin:2004ut,Gualtieri:2003dx} showed the need to understand better SU(3)$\times$SU(3) manifolds and in particular SU(2) manifolds. Some early works \cite{Gauntlett:2003cy,Dall'Agata:2003ir,Dall'Agata:2004dk,Grana:2005sn,Bovy:2005qq,Triendl:2009ap} looked into the matter of compactification on these manifolds. The Ricci scalar of the internal manifold can give precious insights on a particular solution, for example on the presence of sources. Thus, a better understanding of the Ricci scalar is important and in particular its expression in terms of the SU(2) torsion classes in six dimensions (it has already been done for five dimensional manifolds \cite{2007math2790B}).

 This work has been inspired by \cite{2007JGP} where the authors give the expression of the Ricci scalar in terms of the SU(3) structure torsion classes on six dimensional manifold. Then we do something similar in spirit to what was done in \cite{2002math2282C} namely using the SU(3) result to derive the SU(2) result.

In section \ref{SU3SU2}, we review briefly six dimensional manifolds with SU(3) and SU(2) structure. After expressing the SU(3) torsion classes in terms of the SU(2) torsion classes in section \ref{SU2toSU3}, we use one of the formulas of \cite{2007JGP} to express the Ricci scalar in section \ref{result}.

\section{Review of SU(3) and SU(2) structures}\label{SU3SU2}
\subsection{SU(3) structure}
A six dimensional manifold is said to be of SU(3) structure if it admits a globally defined real two-form $J$ and a globally defined complex decomposable three-form $\Omega$ verifying :
\begin{align}
J\wedge\Omega=&0 & J\wedge J\wedge J=&\frac{3i}{4}\Omega\wedge\overline{\Omega}
\end{align}
We can quantify the failure of these forms to be closed using the following torsion classes : 
\begin{align}
 dJ=&3/2 \Im(\bar{W}_1 \Omega)+W_4\wedge J+ W_3\label{torsionSU31}\\
 d\Omega=&W_1 J^2+W_2\wedge J+\bar{W}_5\wedge \Omega\label{torsionSU32}
\end{align}
with $W_1$ a complex scalar, $W_2$ a complex primitive (ie. $J\wedge J\wedge W_2=0$) (1,1) form, $W_3$ a real primitive (ie. $J\wedge W_3=0$) (2,1)+(1,2) form, $W_4$ a real one form and $W_5$ a complex (1,0) form. Constraints on these torsion classes put constraints on the manifold itself. For example, if all the torsion classes vanish, the manifold is Calabi-Yau. If $W_1=W_2=0$, then the manifold is complex etc.
\subsection{SU(2) structure}
A six-dimensional manifold is said to be of SU(2) structure if it admits a globally defined complex one-form $K=K^1+i K^2$, a real two form $j$ and a complex two form $\omega$ verifying :
\begin{align}
\iota_K K=&0 & \iota_{\overline{K}}K=&2\\
j\wedge \omega =&0 & \iota_K j=\iota_K \omega=&0 & j\wedge j=\frac{1}{2}\omega \wedge \overline{\omega}
\end{align}
One can then define a triplet of real two forms $J^a$ with $J^1=j$, $J^2=\Re(\omega)$ and $J^3=\Im(\omega)$. This triplet verifies :
\begin{align}
\iota_K K=&0 & \iota_{\overline{K}}K=&2\\
 \iota_K J^a=&0 & J^a\wedge J^b=&\delta^a_b J^1\wedge J^1
\end{align}

As in the SU(3) structure case, one can define torsion classes as follows\footnote{stricto sensu, the torsion classes should be defined with respect to $j$ and $\omega$ rather than the $J^a$ but the expressions are simpler and more "covariant" using the $J^a$} :
\begin{align}
 dK^i=&m^i_{~a}J^a+m^i_{~0}K^1\wedge K^2+\tilde{\mu}^i_{~j}\wedge K^j+\mu^i\label{torsionSU21}\\
 dJ^a=&n^0_{~i} K^i\wedge J^a+\epsilon^{abc}n^b_{~i} K^i\wedge J^c+\nu^a_{~i}\wedge K^i+\tilde{\nu}_{3}^a+\tilde{\nu}_1^a\wedge K^1\wedge K^2\label{torsionSU22}
\end{align}
$m^i_{~a},~m^i_{~0},~n^a_{~i},~n^0_{~i}$ are 16 real scalars. $\tilde{\mu}^i_{~j}$ are 4 real one-forms verifying $\iota_{K^k} \tilde{\mu}^i_{~j}=0$. $\mu^i,~\nu^a_{~i}$ are 8 real two-forms verifying : $\mu^i\wedge J^b=\nu^a_{~i}\wedge J^b=0$ and $\iota_{K^k} \mu^i=\iota_{K^k} \nu^a_{~i}=0$. $\tilde{\nu}^a_3$ are 3 real three-forms verifying $\iota_{K^k} \tilde{\nu}^a_3=0$ (and thus $\tilde{\nu}^a_3\wedge J^b=0$). Finally $\tilde{\nu}^a_{1}$ are three real one-forms verifying $\iota_{K^k} \tilde{\nu}^a_1$ but they are not independant from one another. Indeed, from considering $d(J^a\wedge J^b=\delta^a_b J^1\wedge J^1)$, one can see that $\tilde{\nu}^2_1$ and $\tilde{\nu}^3_1$ can be expressed in terms of $\tilde{\nu}^1_1$. Nevertheless we will continue to use the three $\tilde{\nu}^a_1$ in order to simplify the expressions. 

To summarize the independant real torsion classes are : 16 scalars, 5 one-forms, 8 two-forms and 3 three-forms.

\section{Going from SU(2) to SU(3)}\label{SU2toSU3}

From a SU(2) structure, one can define a family of SU(3) structure. Let R be a SO(3) matrix and define\footnote{Note that $j_c$ and $\omega_c$ together with $K$ define a SU(2) structure.} $j_c=R^1_{~a}J^a$ and $\omega_c=(R^2_{~a}+i R^3_{~a})J^a$. Then $\hat{J}=j_c+K^1\wedge K^2$ and $\hat{\Omega}=K\wedge \omega_c$ define a SU(3) structure. By computing $d\hat{J}$ and $d\hat{\Omega}$ using (\ref{torsionSU21}) and (\ref{torsionSU22}) and comparing with (\ref{torsionSU31}) and (\ref{torsionSU32}), one obtains the expressions of the SU(3) torsion classes in terms of the SU(2) torsions classes :

\begin{align}
W_1=&\frac{1}{3}\sum_a C_a(m_a^++n^a_+)\label{Wexp1}\\
W_2=&\frac{1}{2}\sum_aC_a\left[2(2m_a^+-n^a_+)(\frac{j_c-2K^1\wedge K^2}{3})+2i\nu^a_++\ast(\tilde{\nu}^a_3\wedge K)-\hat{\nu}^a\wedge K\right]\nonumber\\
&+\frac{1}{2}\sum_{i,j}(\delta^i_j-i\epsilon_{ij})\ast(\tilde{\mu}^+_i\wedge K^j\wedge \omega_c)\label{Wexp2}
\end{align}
\begin{align}
W_3=&\frac{1}{2}\sum_{a,b,c,j}\epsilon_{abc}(m^j_{~b}-n^b_{~j})R^1_{~c} K^j\wedge J^a-\frac{1}{2}\sum_{a,b,j,k}\epsilon_{jk}(R^1_{~a} R^1_{~b}-\delta_{ab})(m^j_{~b}-n^b_{~j})K^k\wedge J^a\nonumber\\
&-\sum_{j,k}\epsilon_{jk}K^j\wedge \mu^k+\sum_{j,a}R^1_{~a} K^j\wedge \nu^a_{~j}-\frac{1}{6}\sum_a \tilde{\nu}_1^a\wedge J^a+\frac{1}{2}\sum_a R^1_{~a} K^1\wedge K^2\wedge \tilde{\nu}_1^a \nonumber\\
&+\frac{1}{2} (K^1\wedge K^2-j_c) \wedge \sum_i\tilde{\mu}^i_{~i}+\frac{1}{2}\sum_a R^1_{~a}\tilde{\nu}_3^a+\frac{1}{2}\sum_{a}R^1_{~a}\hat{\nu}^a\wedge K^1\wedge K^2\label{Wexp3}\\
W_4=&\sum_j n^0_{~j} K^j+\sum_{j,k,a}\epsilon_{jk}m^j_{~a}R^1_{~a} K^k+\frac{1}{2}\sum_a R^1_{~a} \tilde{\nu}_1^a+\frac{1}{2}\sum_i \tilde{\mu}^i_{~i}-\frac{1}{2}\sum_{a}R^1_{~a}\hat{\nu}^a\label{Wexp4}\\
W_5=&\frac{K}{2}(n^0_-+i m^-_0-i R^1_{~a} n^a_-)+\frac{1}{4}\sum_{i,j}(\delta^i_j+i\epsilon_{ij})\left[\tilde{\mu}^i_{~j}+i\ast\left(\tilde{\mu}^i_{~j}\wedge K^1\wedge K^2\wedge j_c \right)\right]\nonumber\\
&-\frac{1}{4}\sum_{a}\overline{C_a}\ast\left[\left(\ast\tilde{\nu}_3^a\right)\wedge \omega_c\right]\label{Wexp5}
\end{align}

with  $C_a=R^2_{~a}+i R^3_{~a}$, $\phi_1+i \phi_2=\phi_+$, $\phi_1-i\phi_2=\phi_-$, $\hat{\nu}^a=\ast\left[(\ast \tilde{\nu}^a_3)\wedge j_c\right]$.

\section{Ricci scalar}\label{result}

In \cite{2007JGP}, the authors gave the expression of the Ricci scalar in terms of the SU(3) torsion classes\footnote{For two forms, $\omega_1$ and $\omega_2$, one defines $<\omega_1,\omega_2>=\ast[\ast(\overline{\omega_1})\wedge \omega_2]$ and $|\omega_1|^2=<\omega_1,\omega_1>$, the codifferential is defined as $\delta \omega_1=\ast(d(\ast \omega_1))$} :
\begin{align}
 R_6=&\frac{15}{2}|W_1|^2-\frac{1}{2}|W_2|^2-\frac{1}{2}|W_3|^2-|W_4|^2+4\delta W_{5R}+2\delta W_4+8<W_4,W_{5R}>
\end{align}

Thanks to our expressions of the SU(3) torsion classes in terms of the SU(2) torsion classes (\ref{Wexp1})-(\ref{Wexp5}), we can give the Ricci scalar in terms of the SU(2) torsion classes :
\begin{align}
 R_6=&-\sum_{a,i}(m^i_{~a})^2+3\sum_i(n^0_{~i})^2+4\sum_{i,j}\epsilon_{ij}n^0_{~i}m^j_{~0}+4\sum_{a,i}m^i_{~a}n^a_{~i}-\frac{1}{2}\sum_i|\mu^i|^2-\frac{1}{2}\sum_{a,i}|\nu^a_{~i}|^2\label{R6expression}\\
 &-\frac{1}{2}\sum_{i,j,k,l}(\epsilon_{ik}\epsilon_{jl}+\epsilon_{il}\epsilon_{jk})\ast(\tilde{\mu}^i_{~j}\wedge\ast(\tilde{\mu}^k_{~l}))-\frac{1}{6}\sum_{a}|\tilde{\nu}_1^a|^2-\frac{3}{2}\sum_{a}|\tilde{\nu}_3^a|^2-\sum_{a,i}\ast\left[\tilde{\mu}^i_{~i}\wedge (\ast \tilde{\nu}_3^a)\wedge J^a\right]\nonumber\\
 &+\frac{1}{3}\sum_{i,j,a}\ast\left[(\ast\tilde{\mu}^i_{~j})\wedge \ast\left(\tilde{\nu}_1^a\wedge J^a\wedge K^i\wedge K^j\right)\right]+\frac{1}{2}\sum_{a,b}(1-\delta^a_b)\ast\left[\ast\left((\ast \tilde{\nu}_3^a)\wedge J^a\right)\wedge(\ast \tilde{\nu}^b_3)\wedge J^b\right] \nonumber\\
 &-2 \sum_{i,j}\epsilon_{ij}\delta(m^i_{~0} K^j)+4\sum_{i}\delta(n^0_{~i} K^i)+2\sum_i \delta(\tilde{\mu}^i_{~i})-\sum_{a}\ast\left((\delta\tilde{\nu}_3^a)\wedge J^a\wedge K^1\wedge K^2\right)\nonumber\\
 &+\sum_{a}R^1_{~a}\left[\delta(\tilde{\nu}_1^a)-\sum_{b,c,i,j}\epsilon_{abc}\epsilon_{ij}(2m^i_{~b}+n^b_{~i})n^c_{~j}+\sum_{i,j}\epsilon_{ij}(2m^i_{~a}+4n^a_{~i})n^0_{~j}-4\sum_{i} m^i_{~a}m^i_{~0}\right.\nonumber\\
&\hspace{1.8cm}+\frac{1}{2}\sum_{b,c,i,j}\epsilon_{abc}\epsilon_{ij}\ast(\nu^b_{~i} \wedge \ast \nu^c_{~j})+\sum_{i,j}\epsilon_{ij}\ast(\nu^a_{~i}\wedge \ast \mu^j)+\sum_{i,j}\ast(\tilde{\mu}^i_{~j}\wedge\tilde{\nu}_3^a\wedge K^i\wedge K^j)\nonumber\\
&\left.\hspace{1.8cm}-\sum_{i,j,k} \ast\left[\tilde{\mu}^i_{~j}\wedge \tilde{\mu}^i_{~k}\wedge K^j\wedge K^k\wedge J^a\right]+2\sum_{i,j}\epsilon_{ij}\delta((m^i_{~a}+n^a_{~i}) K^j)\right.\nonumber\\
&\left.\hspace{1.8cm}-\sum_{b}\ast\left[\left(\tilde{\nu}_1^a\wedge (\ast \tilde{\nu}_3^b)-\frac{1}{3}\tilde{\nu}_1^b\wedge (\ast \tilde{\nu}_3^a)\right)\wedge J^b\right]-\sum_{i,j}\delta[\ast(\tilde{\mu}^i_{~j}\wedge K^i\wedge K^j\wedge J^a)]\right]\nonumber
\end{align}

The curvature of the manifold doesn't depend on the specific SU(2) structure we choose and so shouldn't depend on the matrix $R$. To solve this puzzle, one has to remember that nowhere we requested that $d^2=0$ on the forms defining the SU(2) structure. And indeed, one can show : 

\begin{align}
R_{6|R^1_{~a}~\mathrm{part}}=\ast\sum_{a}R^1_{~a}\left[\frac{1}{2}\sum_{b,c}\epsilon_{a,b,c}d(d(J^b))\wedge J^c-\sum_id(d(K^i))\wedge K^i\wedge J^a\right]=0
\end{align}

so that the Ricci scalar is independent of the matrix $R$.

\section*{Conclusion}
Using the link between SU(3) and SU(2) structures, we were able to express the Ricci scalar in terms of the SU(2) structure torsion classes. Unfortunately, applying the same technique to the Ricci tensor has been unfruitful so far but is still under investigation. Another way to derive this formula could have been to use the result of \cite{Lust:2008zd} where they give the expression of the Ricci scalar in term of the pure spinors\footnote{The author thanks David Andriot for bringing this formula to his attention}. Indeed, one could have expressed the pure spinors in terms of the SU(2)-forms and derived the result.
 
\section*{Acknowledgement}

The author would like to thank Hagen Triendl who participated in the project at an earlier stage. The author is supported in part by INFN and by the European Research Council under the European Union's Seventh Framework Program (FP/2007-2013) ERC Grant Agreement n.307286 (XD-STRING).

\bibliographystyle{utphysmodb}
\bibliography{draftSU2bib}

\end{document}